\documentclass[aps,prl,twocolumn,longbibliography,superscriptaddress]{revtex4-1}

\usepackage{amssymb}
\usepackage{amsbsy}
\usepackage{amsmath}
\usepackage{graphicx}
\usepackage{graphics}
\usepackage{setspace}
\usepackage{array}
\usepackage{color}
\usepackage{fontenc}
\usepackage{textcomp}
\usepackage{bm}
\usepackage{float}
\usepackage[bookmarks=false,linkcolor=blue,urlcolor=blue,colorlinks,citecolor=blue]{hyperref}

\DeclareMathOperator{\im}{Im}

\DeclareMathOperator{\Tr}{Tr}
\DeclareMathOperator{\sgn}{sgn}

\newcommand{\vex}[1]{\bm{\mathrm{#1}}}

\newcommand{\bsub}{\begin{subequations}}
\newcommand{\esub}{\end{subequations}}

\graphicspath{{../Figures/}}

\begin{document}
\title{Non-Hermitian Higher-Order Weyl Semimetals}
\author{Sayed Ali Akbar Ghorashi}\email{sghorashi@wm.edu}
\affiliation{Department of Physics, William $\&$ Mary, Williamsburg, Virginia 23187, USA}
\author{Tianhe Li}
\affiliation{Department of Physics and Institute for Condensed Matter Theory,  University of Illinois at Urbana-Champaign, IL 61801, USA}
\author{Masatoshi Sato}
\affiliation{Yukawa Institute for Theoretical Physics, Kyoto University, Kyoto 606-8502, Japan}

\date{\today}

\newcommand{\be}{\begin{equation}}
\newcommand{\ee}{\end{equation}}
\newcommand{\bea}{\begin{eqnarray}}
\newcommand{\eea}{\end{eqnarray}}
\newcommand{\h}{\hspace{0.30 cm}}
\newcommand{\vs}{\vspace{0.30 cm}}
\newcommand{\n}{\nonumber}

\begin{abstract}
We study non-Hermitian higher-order Weyl semimetals (NHHOWSMs) possessing real spectra and having inversion $\mathcal{I}$ ($\mathcal{I}$-NHHOWSM) or time-reversal symmetry $\mathcal{T}$ ($\mathcal{T}$-NHHOWSM). When the reality of bulk spectra is lost, the NHHOWSMs exhibit various configurations of surface Fermi Arcs (FAs) and Exceptional Fermi Rings (EFRs), providing a setup to investigate them on an equal footing. The EFRs only appear in the region between 2nd-order WNs. We also discover Weyl nodes originating from non-Hermicity, called \emph{non-Hermitian Weyl nodes (NHWNs)}. Remarkably, we find T-NHHOWSMs which host only 2nd-order NHWNs, having both surface and hinge FAs protected by the quantized biorthogonal Chern number and quadrupole moment, respectively. We call this intrinsically non-Hermitian phase \emph{exceptional HOWSM}. In contrast to ordinary WNs, the NHWNs can instantly deform to line nodes, forming a \emph{monopole comet}. The NHWNs also show \emph{exceptional tilt-rigidity}, which is a strong resistance towards titling due to attachment to exceptional structures. This phenomenon can be a promising experimental knob. Finally, we reveal the exceptional stability of FAs called \emph{exceptional helicity}. Surface FAs having opposite chirality can live on the same surface without gapping out each other due to the complex nature of the spectrum. Our work motivates an immediate experimental realization of NHHOWSMs.
\end{abstract}
\maketitle
\emph{Introduction}.--Non-Hermitian and higher-order topological phases are two new branches of topological phases which have ignited numerous attentions over the last few years \cite{ReviewNHRMP,reviewNHUeda,Benalcazar2017-1,Benalcazar2017-2,Song2017,SatoPRX2018,SatoPRX2019,chernLR,BiorthogonalPRL2018,Kawabata_2019,PhysRevLett.124.056802,PhysRevLett.120.146402,NoriPRL2019,SatoskinPRL2020,YaoWangPRL2018,TaylorNHPRB2011,NHclassZhouPRB2019,Langbehn2017,Schindler2018-1,NHCherninsulator,GHHRHOTSC,GHRvortex,CAlugAru2018,tianhe1,PhysRevLett.123.266802,BitanPRR2019,Peterson2018,Noh2018,Serra-Garcia2018,Imhof2018,xue2019acoustic}. Recently, the interplay of these two have also been studied \cite{2ndorderskinSatoPRB2020,Nori2ndorderPRL2019,BiorthHOPRB2019,2ndorderskinFuPRB2021,2ndorderskinOkugawaPRB2020,PhysRevLett.122.195501,lopez2019multiple, PhysRevB.102.094305,PhysRevB.103.L041115,PhysRevLett.123.073601,PhysRevB.102.104109,ZHOU2020125653,hybridskintop,EzawaHONHPRB2019}. However, despite attempts to understand the role of non-Hermiticity in systems having higher-order topology the effect of non-Hermitian perturbation has yet remains to be explored in 3d higher-order semimetals. Very recently we have studied the non-Hermitian higher-order Dirac semimetals (NHHODSMs) \cite{GhorashiNHHodsm2021}. \\
\begin{figure}[t!]
    \centering
    \includegraphics[width=0.45\textwidth]{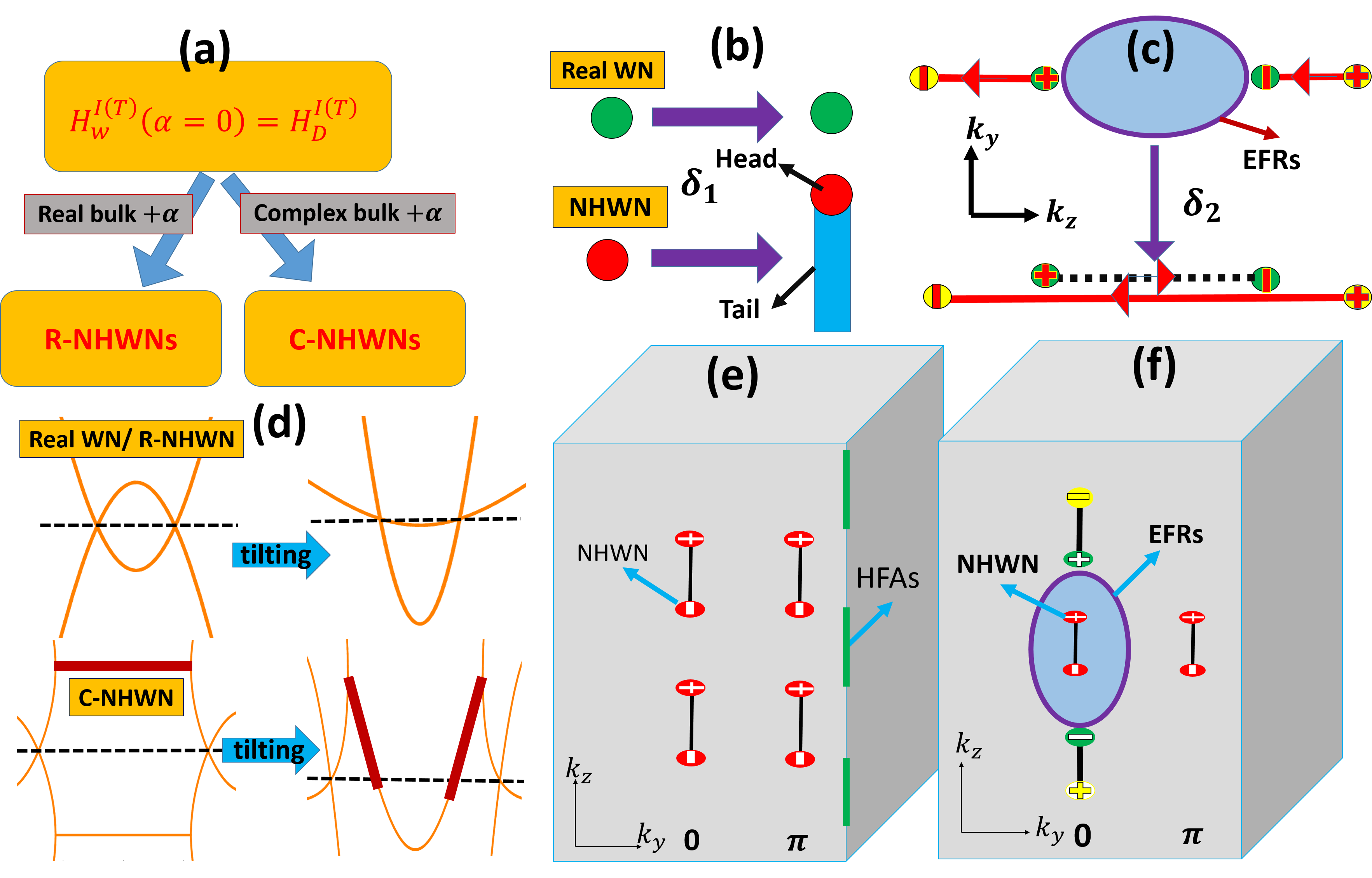}
    \caption{Summary of main results.(a) Upon the addition of $\alpha$ to non-Hermitian Dirac semimetallic phase of \cite{GhorashiNHHodsm2021}( $H^{\mathcal{I}(\mathcal{T})}_{w}(\alpha=0)$) having real (complex) spectrum the R-NHWNs (C-NHWNs) can appear aside from the real WNs of Hermitian limit. (b) An example of instability of NHWNs in the bulk: in presence of the perturbation $\delta_1$ they deform to a intrinsically non-Hermitian nodal line structure, \emph{monopole comet}, that while maintaining their original monopole charge, they are connected on the surface only via the original position of un-deformed NHWNs (head of comet) [see Fig.~\ref{fig:wETRSI}(c) for the surface plot]. (c) The surface instability of EFRs: in presence of the perturbation $\delta_2$ they deform by changing band connectivity of surface FAs generating a region of counter-propagating FAs a phenomenon that we dubbed \emph{exceptional-helicity}.[WNs shifted for visibility] (d) \emph{Exceptional tilt-rigidity}: The C-NHWNs show strong resistance towards tilting due to ESs (red bars) in compare to real WNs and R-NHWNs. (e) \emph{Exceptional NHHOWSMs}: A genuinely non-Hermitian class of $\mathcal{T}$-NHHOWSMs which host only NHWNs of higher-order topology i.e, all the nodes are connected both through the surface \emph{and} hinge. (f) An example of surface configuration in $H^{\mathcal{I}}$, where EFRs and surface FAs overlap. Similar configurations are possible for $H^{\mathcal{T}}$, while it can host HFAs as well. For more surface configurations see Figs.~\ref{fig:I-NHHOWSM},\ref{fig:T-NHHOWSM}. }
    \label{fig:adpic}
\end{figure}
In this letter, we consider the the effect of $C_4$-symmetric non-Hermitian perturbations on the higher-order Weyl semimetals (HOWSMs) \cite{Ghorashihowsm2020}. We find $\mathcal{I}$ and $\mathcal{T}$ symmetric models that possess real bulk spectrum up to a critical strength of non-Hermiticity while respecting anti-PT symmetry. We show that, when the complexity emerge, these models can host a novel classes of non-Hermitian Weyl nodes (NHWNs) that are intrinsic to Non-Hermitian systems. We characterize these new WNs using biorthogonal Chern number and a new 1D winding number. Both $\mathcal{I},\mathcal{T}$-symmetric models feature various configurations of coexisting real and NHWNs in bulk as well as surface Fermi Arcs (FAs) and exceptional Fermi Rings (EFRs) that are protected by symmetries of class $\mathcal{P}$AI \cite{NHclassSatoPRL2019}.
\begin{figure*}[htb!]
    \centering
    \includegraphics[width=0.95\textwidth]{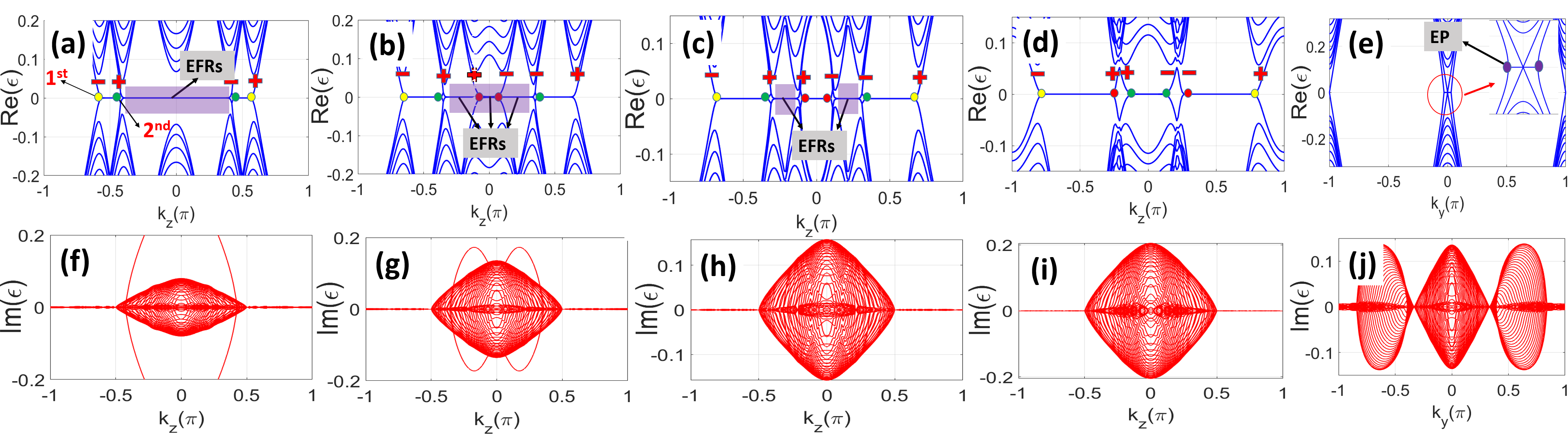}
    \caption{(a-d) The real part of $x$-surface along $k_z (k_y=0)$ for $H^{\mathcal{I}}$ having $\gamma=-1,m_1=0.78$ and $\alpha=0.2,0.35,0.42,0.6$, respectively. (e) The corresponding cut along $k_y(k_z=0)$ for (b). (f-j) The imaginary counterparts of (a-e). The yellow, green, red dots and purple shade denote the $1st$, $2nd$-order, NHWNs and EFRs, respectively. See \cite{sm} for plots of their corresponding $k_y=\pi$ cut and bulk spectrum.}
    \label{fig:I-NHHOWSM}
\end{figure*}
Interestingly, the $\mathcal{T}$ symmetric model additionally host real hinge FAs having quantized biorthogonal quadrupole moments exhibiting a unique Non-Hermtian higher-order Weyl semimetal (NHHOWSMs). Strikingly, we find that by gaping out all the real WNs, a novel and genuinely non-Hermitian Weyl phase, dubbed as \emph{exceptional HOWSM}, emerges which hosts only $2nd$-order NHWNs. Furthermore, we reveal some distinct properties of NHWNs. First, we show that unlike the real WNs they can be destabilized in presence of a Hermitian perturbations such that they can instantly deform to a novel nodal line structure which we dubbed, \emph{monopole comet}, where while maintaining their monopole charges, they connect on the surface FAs at original position of NHWNs (head of comet).
Second, interestingly, we show that some classes of NHWNs exhibit strong resistance towards tilting due to presence of ESs in compare to their real counterparts. A phenomenon that we call \emph{exceptional tilt-rigidity} and can be used as a promising experimental probe/knob. Finally, we reveal a remarkable phenomenon that we call it \emph{exceptional helicity} in which surface band connectivity can be altered while removing EFRs in a way that FAs having opposite chilrality live in the same surface without gapping out each other. Fig.~\ref{fig:adpic} summarizes some of the main results of our work.

\emph{$H^{\mathcal{I}}_w$ Model}.---We start by the following $\mathcal{I}$-symmetric non-Hermitian model,
\begin{align}\label{I-NHHOWSM}
    H^{\mathcal{I}}_{w}=H^{\mathcal{I}}_{HODSM}(\vex{k}) + im_1\Gamma_0a_0(\vex{k}) + \alpha\sigma_0\kappa_2
\end{align}
where $H_{HODSM}(\vex{k})=\sum^{4}_{i=1} a_i(\vex{k})\Gamma_i$ \cite{Lin2017}, $a_4(\vex{k})=\left(\gamma +\frac{1}{2}\cos k_z + \cos k_x\right),\,a_3(\vex{k})=\sin(k_x),\,a_2(\vex{k})=\left(\gamma+\frac{1}{2}\cos k_z+\cos k_y\right),\, a_1(\vex{k})=\sin(k_y)$, and $\gamma$ represent the intra-cell coupling.
$\{\Gamma_\alpha\}$ are direct products of Pauli matrices, $\sigma_i,\kappa_i$, following $\Gamma_0=\sigma^3\kappa^0,\Gamma_i=-\sigma^2\kappa^i\,\textrm{for}\,i=1,2,3,$ and $\Gamma_4=\sigma^1\kappa^0$, $a_0(\vex{k})=\left(\cos(k_x)-\cos(k_y)\right)$ and $m_1$ is a real constant. The amplitudes of inter-cell hoppings is set to $1,$ and we will work in the $C_4^z$ symmetric limit. For $\alpha=0$ but $m_1\neq 0$ the Eq.~\eqref{I-NHHOWSM} represent a NHDSM phase which shows real anti-PT symmetric bulk hosting Dirac nodes while the surface is PT symmetric and complex. The surface become complex by developing EFRs connecting the projections of two Dirac nodes on the surface and as a result the original hinge FAs of $H_{HODSM}$ gap out. On the oter hand, for $\alpha\neq 0,\,m_1=0$, $H^{I}_w$ is a $\mathcal{I}$-HOWSM which can host a pair of $1st$-order or $2nd$-order WNs or combination of both types of WNs \cite{Ghorashihowsm2020}. The $2nd$-order WNs are defined as the transition point between a phase having non-zero Chern number to the one having quantized quadrupole moment ($q_{xy}$). Therefore, they show coexistence of surface and hinge FAs. When both $m_1$ and $\alpha$ are non-zero, the bulk is still anti-PT symmetric but chiral symmetry and combination of reciprocity and $\mathcal{I}$ are broken, hence $H^{\mathcal{I}}_w$ belongs to the class $\mathcal{P}$C$^{\dagger}$ of non-Hermitian topological phases\cite{NHclassSatoPRL2019}.\\
Similar to non-Hermitian Dirac semimetallic phases in \cite{GhorashiNHHodsm2021}, interestingly, the bulk spectrum remains real up to a critical value of $m_1$. The spectrum of Eq.~\eqref{I-NHHOWSM} can be obtained as $E_{\pm,\pm} (\vex{k})= \pm \sqrt{f(k)+\alpha^2 \pm 2\alpha \sqrt{g(k)}}$, where $f(k)=\sum^4_i a^2_i(\vex{k})-m_1^2a^2_0(\vex{k})$ and $g(k)=a^2_2(\vex{k})+a^2_4(\vex{k})-m_1^2a^2_0(\vex{k})$. At $(k_x, k_y) = (0, \pi), (\pi, 0)$, $a_1=a_3=0$, interestingly, when $\sqrt{a^2_2(\vex{k})+a^2_4(\vex{k})-m_1^2a^2_0(\vex{k})}-|\alpha|=0$ a new set of WNs may emerge coexisting with real WNs and are located at $\cos(k_z)=-2\gamma\pm \sqrt{2}\sqrt{4m_1^2+\alpha^2-2}$. This is particularly interesting as heuristically addition of non-Hermitian perturbations to Weyl/Dirac nodes results in deformation of the nodes to exceptional rings\cite{ReviewNHRMP,NHWSM1}. Here, on the other hand, NHWNs are indeed point-like nodes. We call these new nodes \emph{Non-Hermitian Weyl nodes (NHWNs)} and we will show that while they share some common properties with real WNs, they behave distinctly in presence of an external Hermitian perturbations. Interestingly, NHWNs can appear in two types. If we start with a NHDSM phase having real (complex) spectra, the emergent NHWNs can be detached (attached) to bulk exceptional structures (ESs) and henceforth and we denote them as R-NHWNS (C-NHWNs) \cite{sm}.

\begin{figure*}[htb!]
    \centering
    \includegraphics[width=0.9\textwidth]{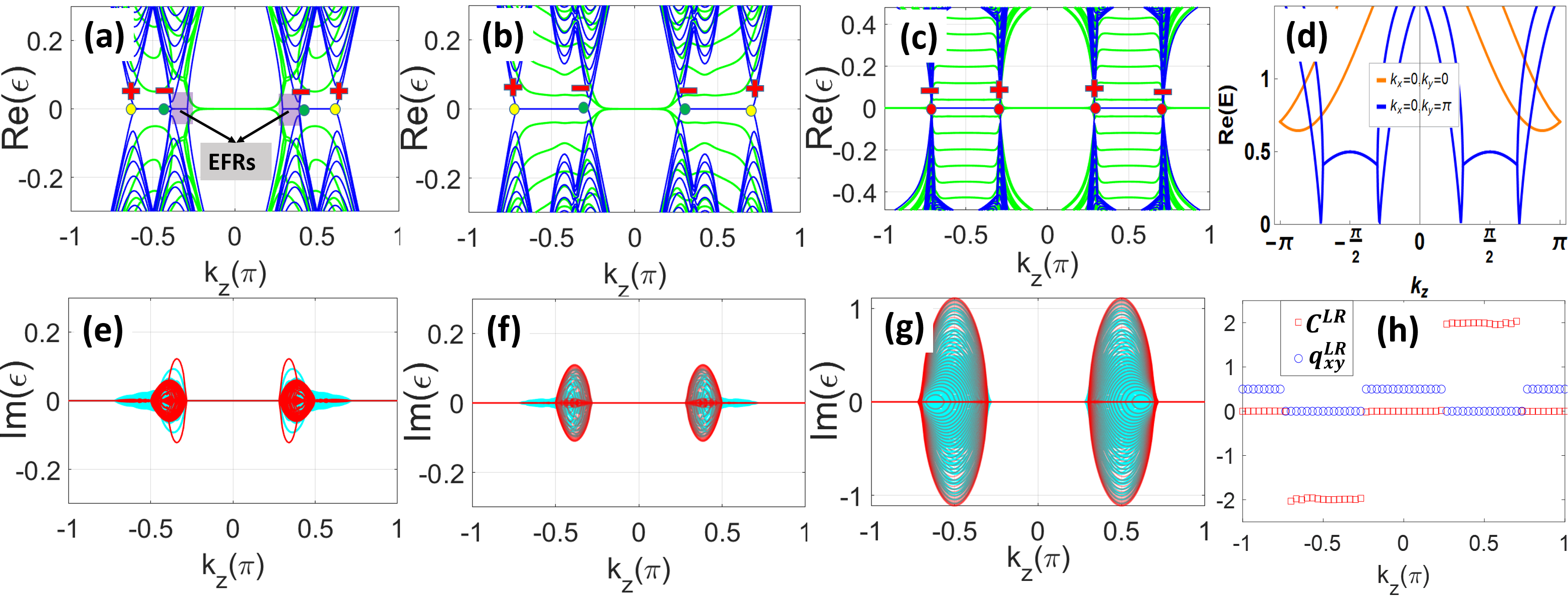}
    \caption{The real part of spectrum at $x$-surface ($k_y=0$) (blue) and $z$-hinge (green) of $H^{\mathcal{T}}$ having (a) $\gamma=-1,m_2=0.85,\alpha=0.3$, (b), $\gamma=-1,m_2=0.85,\alpha=0.6$ (c) $\gamma=0,m_2=0.9,\alpha=0.5$. (d) The real part of bulk spectrum for (c,g) along $k_z$ for $k_x=k_y=0$(orange) and $k_x=0,k_y=\pi$ (blue). (e-g) The imaginary counterparts of (a-c)[red:$x$-surface, cyan: $z$-hinge]. (h) The corresponding biorthogonal chern number ($\mathcal{C}^{LR}$, red) and quadrupole momment ($q^{LR}_{xy}$, blue) of (c,d,g). }
    \label{fig:T-NHHOWSM}
\end{figure*}

\begin{figure}[b!]
    \centering
    \includegraphics[width=0.46\textwidth]{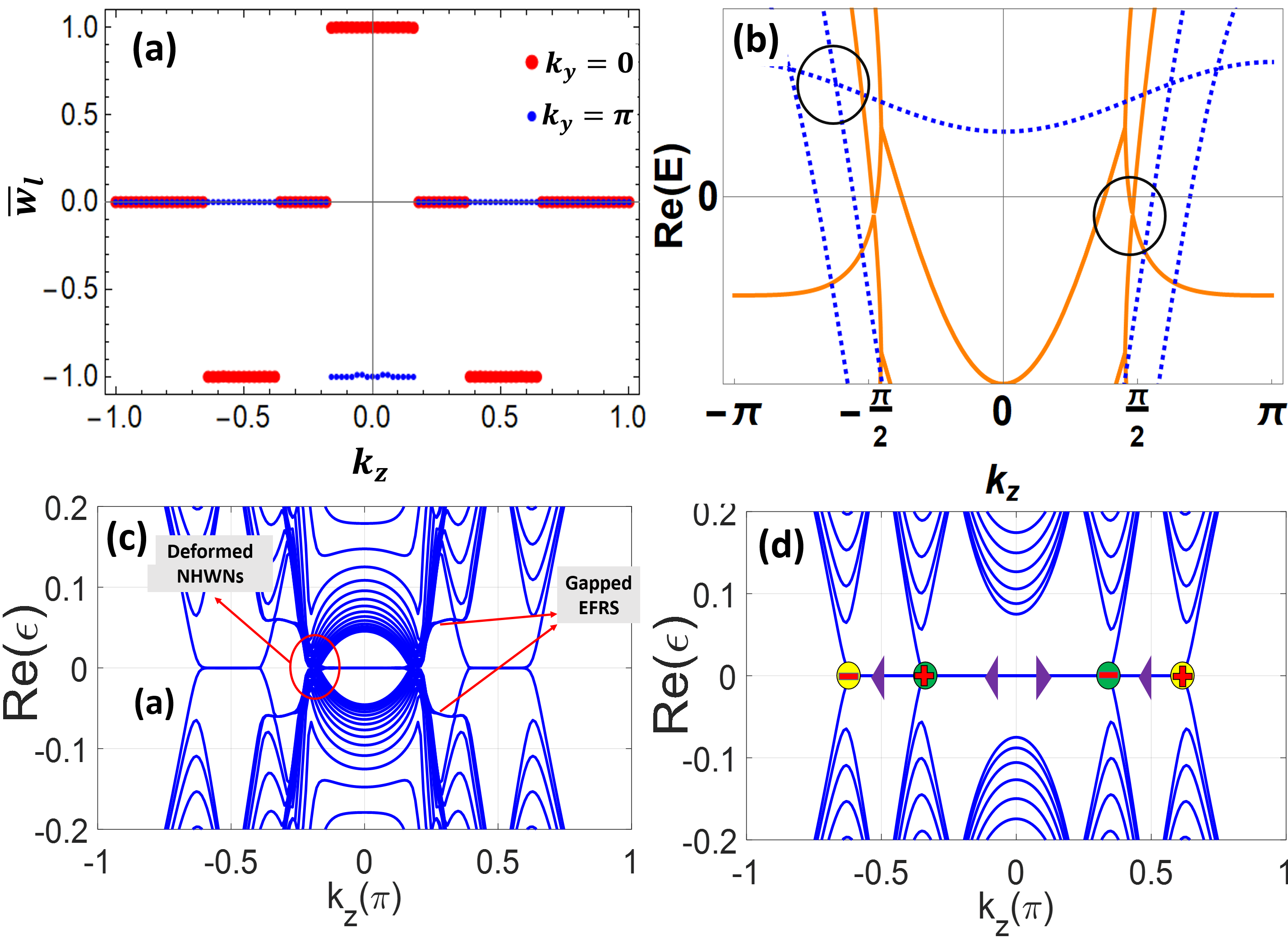}
    \caption{(a) The winding number $\bar{w_l}$ at $k_y=0$(red) and $k_y=\pi$(blue) planes for $H^{\mathcal{I}}_w$.[$\gamma=-1,\,m_1=0.8,\alpha=0.3$] (b) The real (dashed blue) and C-NHWNs (solid orange) of $H^{\mathcal{I}}_w$ in presence of tilting perturbation $0.8\sin(k_z-pi/2)\mathbb{I}_4$ [$\gamma=-0.6,\alpha=0.3,m_1=0.8$]. (c) The $x$-surface of $H^{\mathcal{I}}$ in presence of $\delta_1=0.05$ showing the deformed NHWNs and gapped EFRs. (d) $H^{\mathcal{I}}$ in presence of $\delta_2=0.05$, monopole charges and direction of chiralities are denoted by dots and arrows, respectively (see \cite{sm} for plot of $\im(E)$).}
    \label{fig:wETRSI}
\end{figure}

Now let us investigate the surface. Fig.~\ref{fig:I-NHHOWSM}, shows the evolution of $x$-surface states [$y$-surface is identical] as tuning the $\alpha$ for a fixed value of $m_1$ \cite{note3}. By increasing $\alpha$ from zero the real bulk Dirac nodes split. On the surface, interestingly, the EFRs survive only between two $2nd$-order WNs, where in the limit of $m_1=0$ shows $q_{x,y}=0.5$ Fig.\ref{fig:I-NHHOWSM}(a). This is because, phenomenologically, the EFRs emerge by removing hinge FAs, which means that when $H^{\mathcal{I}}$ hosts only $1st$-order WNs the EFRs do not appear, indicating an intimate interplay of non-Hermitian and higher-order topology. The real WNs ($1st$ and $2nd$-order) similar to the Hermitian limit are connected by conventional surface FAs, providing a platform that hosts both the conventional FAs and EFRs at different momenta. \\
When the NHWNs appear, their projections on $x$-surface connect at both $k_y=0,\pi$ and form a region where the EFRs and surface FAs overlap (Fig.~\ref{fig:I-NHHOWSM}(b,e)). Then, as $\alpha$ increases, through series of surface phase transitions, the EFRs gaps out between NHWNs and only survive in the region between the $2nd$-order and NHWNs (note they have same monopole charges)  Fig.~\ref{fig:I-NHHOWSM}(c). By further increasing of $\alpha$ the NHWNs exchange their positions with $2nd$-order WNs, but in doing so, they exchange their band connectivity at $k_y=0$ through a surface phase transition in a way that instead each NHWNs connects to real $1st$-order WNs Fig.~\ref{fig:I-NHHOWSM}(d) and all EFRs are gapped out. \\
The $x$-surface possess PT symmetry while the bulk respects anti-PT symmetry then all the models discussed in this work, present another examples of 3D higher-order \emph{hybrid-PT topological phases} \cite{GhorashiNHHodsm2021}. The topological protection of EFRs can be understood as follows. On the $x$-surface, only the $\mathcal{T}\mathcal{M}_{y}\mathcal{M}_z$ is preserved and hence according to \cite{NHclassSatoPRL2019}, it belongs to class $\mathcal{P}$AI of gapless non-Hermitian topological phases. As a result, the EFRs on the surface are protected by a $\mathbb{Z}_2$ topological number\cite{note2}.\\
\emph{$H^{\mathcal{T}}_w$ Model}.---We consider the following $\mathcal{T}$-symmetric model,
\begin{align}\label{T-NHHOWSM}
    H^{\mathcal{T}}_{w}=H^{\mathcal{T}}_{HODSM}(\vex{k}) + im_2\Gamma_0a_0(\vex{k})\sin(k_z) + \alpha\sigma^0\kappa^2\sin(k_z)
\end{align}
similar to Eq.~\eqref{I-NHHOWSM}, $H^{\mathcal{I}}_w$ belongs to the class $\mathcal{P}$C$^{\dagger}$ and preserves anti-PT, even though it has a broken $\mathcal{I}$ symmetry. For $\alpha=0$, $H^{\mathcal{T}}$ represents a $\mathcal{T}$-NHHODSM. On the other hand, in the Hermitian $m_2=0$ limit, Eq.~\eqref{T-NHHOWSM} is a $\mathcal{T}$-HOWSM. Due to $\mathcal{T}$, the minimum number of all the WNs are four. Therefore, on each of  $(0,\pi)$ and $(\pi,0)$ axes, at some finite vale of $\alpha$ four NHWNs emerge. As a result on the surfaces, there are two separate patches of EFRs separated by a real gap. Here, again, we observe that EFRs can only appear in the region between two real $2nd$-order WNs. Remarkably, by further opening boundary along the $x$ and $y$-directions to get a hinge, we find hinge FAs (HFAs) that survive in a region corresponding to the gapped region of the surface states \ref{fig:T-NHHOWSM}(a,e). However, by further increasing of $\alpha$, the EFRs can be removed and like the Hermitian limit, the HFAs are connected to $2nd$-order \ref{fig:T-NHHOWSM}(b,f). %
%
Strikingly, by removing all the real WNs via decreasing $\gamma$, we obtain a $\mathcal{T}$-NHHOWSM that possess hinge FAs both in the middle and at the end of BZ \ref{fig:T-NHHOWSM}(c,d,g). This means all the NHWNs are of $2nd$-order type and so are connected both through surface and hinges. This is in sharp contrast to the parent Hermitian $\mathcal{T}$-HOWSMs which only two out of four WNs are $2nd$-order. We emphasize that this is a novel HOWSM which is intrinsic only to the NH phases as the NHWNs can only exist in presence of NH perturbation. We call this new phase \emph{exceptional higher-order Weyl semimetal}. \\
In order to characterize the zero-mode at the hinges, we employ a biorthogonal \cite{BiorthogonalPRL2018,note4} real-space formula \cite{Qxyoperator1,Qxyoperator2,Qxyoperator3} for quadrupole moment, $q^{LR}_{xy}$, which is shown to correctly capture the higher-order topology in NHHODSMs \cite{GhorashiNHHodsm2021}. As is evident in the case of exceptional HOWSMs, for the regions having hinge FAs the $q^{LR}=0.5$ (see Fig.~\ref{fig:T-NHHOWSM}(h)).
\emph{Topological charges of WNs}.--- We use two methods to characterize the topological character of WNs. First, we obtain the monopole charge of each WNs by computing the biorthogonal real-space formula of Chern number ($\mathcal{C}^{LR}$) at each 2D $k_z$-slice \cite{chernLR,note1} (e.g., see Fig.~\ref{fig:T-NHHOWSM}(h)). Second, we introduce a new 1D winding number.  Using the fact that each of $k_y=0,\pi$ ($k_x=0,\pi$) planes separately respect the sublattice symmetry (even though the full 3D Hamiltonian of $H^{\mathcal{I},\mathcal{T}}_w$ do not),  $H^{\mathcal{I},\mathcal{T}}_w$ can be block-off diagonalized at $k_x=0,\pi$ ($k_y=0,\pi$) planes having $\mathcal{Q}_{1,2}$ blocks. Note that, in general for non-Hermitian systems $Q_2\neq Q_1^{\dagger}$. Therefore, we define a new $k_z$-dependent 1D winding number as following,
\begin{align}
    w_{l,(1,2)}(k_z)=\int^{\pi}_{-\pi} \frac{1}{2\pi i} \partial_{k_x} \ln{\det [Q_{(1,2)}]}
\end{align}
where $l=k_y=0,\pi$. Interestingly, by tuning the $k_z$ the $w_l$ changes by $\pm 1$ as crossing any WNs. We emphasize that this is general and can detect both the real and (C,R)-NHWNs. The total $w_l=w_{l,1}+w_{l,2}$ are zero. However, the difference $\bar{w_{l}}=(w_{l,2}-w_{l,1})/2$ are non-zero and quantized to $\pm 1$. Fig.~\ref{fig:wETRSI}(a), shows the $\bar{w_l}$ for $k_y=0,\pi$. For the real WNs on the $k_x=0$ cut, $\bar{w_0}$ is consistent with the monopole charges of $\pm$. However, interestingly, the NHWNs at $(0,\pi)$ and $(\pi,0)$ carry opposite winding numbers. Therefore, their sum $w=w_0+w_{\pi}$ is zero but their difference reflects the Chern number $\bar{w}=\bar{w_0}-\bar{w_{\pi}}$.\\
\emph{Stability of NHWNs}.---Let us now investigate the stability of NHWNs introduced in this work. The Hermitian/real WNs and their corresponding surface FAs are stable and only can be gapped or deformed through a topological phase transition. On the other hand, here we show that the NHWNs can be deformed instantly, upon applying an external Hermitian perturbation. We choose a $C_4$-symmetric perturbation,  $\delta_{1}a_0(\vex{k})\sigma^3\kappa^{0}$ as an example and for sake of brevity, we only focus on the inversion symmetric model. Firstly, we note that the $\delta_1$ break the symmetries of $\mathcal{P}$AI and so gap out the EFRs. More importantly, upon application of $\delta_1$ in the bulk the NHWNs deform to a unique nodal line we call \emph{monopole comet} that preserves the same monopole charge of original NHWNs in Fig.~\ref{fig:wETRSI}(c), but on the surface their corresponding FAs are connected at the original position of NHWNs (i.e, head of comet). This distinct behavior of NHWNs can be used to distinguish them from real stable Hermitian WNs \cite{note5}.

%
\emph{Exceptional Tilt-Rigidity}.--- Type-II WSMs have dispersion that is strongly anisotropic around the Weyl nodes such that its slope changes sign along some directions \cite{reviewweyl}. Interestingly, we find that the C-NHWNs show strong resistance against tilting due to presence of ESs, a phenomenon which we call \emph{Exceptional tilt-rigidity}. Fig.~\ref{fig:wETRSI}(b), shows the real WNs at $(k_x=0,k_y=0)$ (dashed blue) and NHWNs at $(k_x=0,k_y=\pi)$ (orange) in presence of a tilting perturbation. As is evident for the same amount of tilt strength the C-NHWNs are tilted much lesser compare to the case of real WNs. This is true also in compare to the R-NHWNs. This simple but interesting phenomenon provide an experimental knob for detecting of NHWNs.

\emph{Exceptional Helicity}.--- Now we reveal the intriguing surface instability of EFRs. In the presence of $\delta_{2}a_0(\vex{k})\sigma^3\kappa^{2}$, remarkably, the EFRs deform and in doing so the band connectivity of the FAs connecting the WNs is altered. In Fig.~\ref{fig:wETRSI}(d), as a result of $\delta_2$ the two WNs at opposite $\vex{k}$ connect to their partner at $-\vex{k}$. This lead to the area on the surface having counter-propagating FAs. It is noteworthy that this can not occur in Hermitian WSMs. In order to understand this we note that at $k_y=0,\pi$ the $x$-surface respects the anti-PT symmetry which then enforces that a band with energy $E$ to be paired with a band with energy $-E^*$. Therefore, the helical FAs with different $\im(E)$ cannot gap out each other in the real part of energy. We refer to this intrinsically non-Hermitian phenomena as $\emph{exceptional helicity}$.\\
\indent\emph{Experimental remarks}.--- The 2d quadrupole insulators, which are the building blocks of HOWSM models of our work, have already been realized both in Hermitian \cite{Peterson2018,Noh2018,Serra-Garcia2018,Imhof2018} and non-Hermitian regimes \cite{Gao_2021,zhang2ndordersonic}. Recently, Hermitian higher-order semimetals have also experimentally realized in multiple platforms \cite{Wei_2021,ni2021higher,luo2021observation}. In particular, Ref.\cite{ni2021higher}, have explicitly implemented the $H^{\mathcal{I}(\mathcal{T})}(m_{1,(2)}=0)$ model. Therefore, our results can readily be realized in various experiments.

\emph{Acknowledgement}.--- We thank Taylor Hughes for useful discussion. S.A.A.G acknowledges support
from ARO (Grant No. W911NF-18-1-0290) and NSF
(Grant No. DMR1455233). T.L. thanks the US Office of Naval Research (ONR) Multidisciplinary University Research Initiative (MURI) grant N00014-20-1-2325 on Robust Photonic Materials. M.S. was supported by JST CREST Grant No. JPMJCR19T2, Japan, and KAKENHI Grant No. JP20H00131 from the JSPS.


\begin{thebibliography}{65}
\bibitem [{not({\natexlab{a}})}]{note3}%
  \BibitemOpen
  \href@noop {} {\bibfield  {journal} {\bibinfo  {journal} {In this case the
  NHWNs which appear are of type R-NHWNs. There are alternatives to
  Fig.~\ref{fig:I-NHHOWSM}. We can instead fix the $\alpha$ and tune $m_1$,
  which then the C-NHWNs would appear. Note that the general trend particularly
  the Fig.~\ref{fig:I-NHHOWSM}(a,b and d) can occur and only details of surface
  phase transitions from b-d could differ.In \cite{sm}, we have shown one
  alternative example. Also, We note that unlike the $\alpha=0$ limit of
  \cite{GhorashiNHHodsm2021} the surface states become complex only when bulk
  is complex as well.}\ } ({\natexlab{a}})}\BibitemShut {NoStop}%
\bibitem [{not({\natexlab{b}})}]{note2}%
  \BibitemOpen
  \href@noop {} {\bibfield  {journal} {\bibinfo  {journal} {Because of
  $\mathcal{T}\mathcal{M}_y\mathcal{M}_z$, one can show that $\det H(k_y,k_z)$
  is real, and it is non-zero except on the exceptional ring. Then $\sgn[\det
  H(k_y,k_z)]$ gives the $\mathbb{Z}_2$ topological number}\ }
  ({\natexlab{b}})}\BibitemShut {NoStop}%
\bibitem [{not({\natexlab{c}})}]{note4}%
  \BibitemOpen
  \href@noop {} {\bibfield  {journal} {\bibinfo  {journal}
  {$q_{x,y}=\,\frac{1}{2\pi}\im\bigg[\ln{\langle{\Psi^{R(L)}|\hat{U}_{x,y}|\Psi^{R(L)}\rangle}}\bigg]$,
  where $\hat{U}_{x,y}=e^{2 i\pi\sum_r r_x r_y \hat{n}_r/(L_x
  L_y)},\,|\Psi^{R(L)}\rangle = \prod_{n \in occ}
  \gamma^{\dagger}_{n,R(L)}|0\rangle$, $\Psi^{R(L)}$ denote the right and left
  eigenvecors and $x$ ($N_x$) and $y$ ($N_y$) are the coordinate operator
  (sample size) along $x$ and $y$ directions, respectively.}\ }
  ({\natexlab{c}})}\BibitemShut {NoStop}%
\bibitem [{not({\natexlab{d}})}]{note1}%
  \BibitemOpen
  \href@noop {} {\bibfield  {journal} {\bibinfo  {journal}
  {$\mathcal{C}=\frac{2\pi
  i}{L'_xL'_y}\Tr'\big(\mathcal{P}_{\alpha}\big[[x,\mathcal{P}_{\alpha}],[y,\mathcal{P}_{\alpha}]\big]\big),\,
  \mathcal{P}_{\alpha}=\sum_{n\in \alpha}|nR\rangle\langle nL|$, $L'_ {x,y}$
  are the bulk size of a fully bounded lattice and the trace $\Tr'$ sums over
  the inner area (i.e. we can remove $l_{x,y}$ unit cells from the edge such
  that $L'_ {x(y)} = L_{x(y)}-l_{x(y)}$)}\ } ({\natexlab{d}})}\BibitemShut
  {NoStop}%
\bibitem [{not({\natexlab{e}})}]{note5}%
  \BibitemOpen
  \href@noop {} {\bibfield  {journal} {\bibinfo  {journal} {The $\delta_1$ gaps
  out the EFRs with opposite signs and then induces hinge domain walls,
  resulting a $\mathcal{I}$-NHHOWSM having surface FAs and hinge states.
  Similarly, one can use the $C_4$-breaking version of these perturbations as
  $\delta'_{1,2}\sigma^3\kappa^{0,2}$ (see \cite{sm}). We did not find a
  significant difference in terms of bulk nodes and surface. However, breaking
  $C_4$ will gap out the hinge FAs in $H^{\mathcal{T}}$.}\ }
  ({\natexlab{e}})}\BibitemShut {NoStop}%
\end{thebibliography}

%

\end{document}